\documentclass[11pt,a4paper]{amsart}

\usepackage[latin1]{inputenc}
\usepackage[english]{babel}
\usepackage{amsmath}

\usepackage{amssymb, mathabx}
\usepackage[numbers]{natbib}
\usepackage{graphicx}
\usepackage{braket}
\usepackage[pdftex,plainpages=false,colorlinks,hyperindex,bookmarksopen,linkcolor=red,citecolor=blue,urlcolor=blue]{hyperref}

\usepackage{mathrsfs}

\bibpunct{[}{]}{;}{n}{,}{,}
\usepackage[hmargin=3cm,vmargin={3.5cm,4cm}]{geometry}

\newcommand{\be}{\begin{eqnarray}}
\newcommand{\ee}{\end{eqnarray}}

\newcommand{\N}{\mathbb{N}} 
\def\eg{{\it e.g. }} 
\def\ie{{\it i.e. }}

\newcommand{\wt}[1]{\widetilde{#1}}

\def\ds{\displaystyle}

\def\ds{\displaystyle}
\def\eg{{\it e.g.}\ }
\def\ie{{\it i.e.}\ }

\numberwithin{equation}{section}

\allowdisplaybreaks

\begin{document}
\title{On the propagation of transient waves in a viscoelastic Bessel medium}

    \author{Ivano Colombaro$^1$}
		\address{${}^1$ Department of Information and Communication Technologies, 
		Universitat Pompeu Fabra. C/Roc Boronat 138, 08018, Barcelona, SPAIN.}
		\email{ivano.colombaro@upf.edu}
	
	    \author{Andrea Giusti$^2$}
		\address{${}^2$ Department of Physics $\&$ Astronomy, University of 	
    	    Bologna and INFN. Via Irnerio 46, 40126, Bologna, ITALY,
    	    and
    	    Arnold Sommerfeld Center, Ludwig-Maximilians-Universit\"at, Theresienstra{\ss}e 37, 
    	    80333, M\"unchen, GERMANY }
		\email{andrea.giusti@bo.infn.it}
	
    \author{Francesco Mainardi$^3$}
    	    \address{${}^3$ Department of Physics $\&$ Astronomy, University of 	
    	    Bologna and INFN. Via Irnerio 46, 40126, Bologna, ITALY.}
			\email{francesco.mainardi@bo.infn.it}


    \thanks{Paper published in \textbf{Z.~Angew.~Math.~Phys. (2017) 68:~62}, \textbf{DOI}: \href{http://link.springer.com/article/10.1007/s00033-017-0808-6}{10.1007/s00033-017-0808-6}.}

    \date  {\today}

\begin{abstract}
	In this paper we discuss the uniaxial propagation of transient waves within a semi-infinite viscoelastic Bessel medium. First, we provide the analytic expression for the response function of the material as we approach the wave-front. To do so, we take profit of a revisited version of the so called Buchen-Mainardi algorithm. Secondly, we provide an analytic expression for the long time behavior of the response function of the material. This result is obtained by means of the Tauberian theorems for the Laplace transform. Finally, we relate the obtained results to a peculiar model for fluid-filled elastic tubes. 
\end{abstract}

    \maketitle

\section{Statement of the problem} \label{Sec-1}
	In this paper we deal with a semi-infinite (\ie $x\geq 0$) homogeneous and isotropic viscoelastic medium of density $\rho$, which is assumed to be at rest for $t < 0$. For $t \geq 0$, the accessible extremum of the body (defined by $x=0$) is subjected to a perturbation denoted by $r_0 (t)$. Here, we are interested in the solution of the wave equation for the response function of the material $r (t, x)$. In the framework of a linear theory, this problem can be analyzed by working in either the creep or the relaxation representation. Indeed, it is known that a linear viscoelastic body is featured by either the creep compliance (the strain response to a step input stress) or the relaxation modulus (the stress response to a step input of strain), for further details see \cite{Christensen, Mainardi_BOOK10, Pipkin}. 
	
	We further assume that the dynamics of the material is described in terms of a given Bessel model of order $\nu$ (see \cite{IC-AG-FM-2016, IC-AG-FM-Conf, AG-FCAA-2017}). For such a model one has that the dynamical equation for the response function $r(t,x)$, in the creep representation, is given by
	\begin{equation} \label{eq-motion}
	\Big[1 + \Psi (t; \nu) \, \ast \, \Big] \, 
	\frac{\partial ^2 r}{\partial t ^2} = 
	c^2 \, \frac{\partial ^2 r}{\partial x ^2} \, , \quad \nu > -1 \, ,
	\end{equation}
where $\ast$ represents the Laplace convolution, $c$ is the \textit{wave-front velocity} and $\Psi (t; \nu)$ is the \textit{rate of creep function} for the model. 

	If we denote by $J (t; \nu)$ the creep compliance, and with $J_0 \equiv J(0^+; \nu)$, then we have that
	\begin{equation}
\Psi(t; \nu) =  \frac{1}{J_0}\,  \frac{dJ (t; \nu)}{dt} \, , 	
\end{equation}
and
\begin{equation}
c = \frac{1}{\sqrt{\rho \, J_0}} \, .
\end{equation}
	
	In particular, for our model the rate of creep, in the Laplace domain, is provided by (see \cite{IC-AG-FM-2016}) 
\begin{equation}
\widetilde{\Psi} (s; \nu) = \frac{2 (\nu + 1)}{\sqrt{s}} \frac{I_{\nu + 1} (\sqrt{s})}{I_{\nu + 2} (\sqrt{s})} \, , 
\end{equation}
so that the rate of creep function, in the time domain, is given by
\begin{equation}
\Psi (t; \nu) = 4 (\nu + 1) (\nu + 2) + 4 (\nu + 1) \sum_{n=1} ^\infty \exp \left( - j_{\nu + 2 , \, n} ^{2} \, t \right) \, , 
\end{equation}
where $I_\nu$'s are the modified Bessel functions of the first kind, $j_{\nu + 2 , \, n}$ is the $n$th positive real zero of the Bessel function of the first kind $J _{\nu + 2}$. 

	It is worth remarking that the rate of creep is a completely monotonic function with a \textit{weak power law singularity} as $t \to 0^+$. 

	Given this brief introduction to the Bessel models of linear viscoelasticity, we have that the aim of this paper is to compute the response $r(t,x)$ of the material for $x > 0$ as $t \to (x/c) ^+$ and $t \to \infty$.


\section{The wave-front expansion} \label{Sec-2}
	
	First of all, let us summarize an important technique that allows us to compute the asymptotic wave-front expansions for \textit{singular} viscoelastic models (\ie models that exibit a non-analytic creep compliance featured by power law singularities as $t \to 0^+$). Such a procedure was first introduced by Buchen and Mainardi in \cite{Buchen-Mainardi 1975}.
	
	The creep representation of the equation of motion (\ref{eq-motion}), in the Laplace domain, is given by
	\begin{equation} \label{eq-motion-laplace}
	\left[ \frac{\partial ^2}{\partial x^2} - \mu ^{2} (s) \right] \wt{r} (s, x) = 0 \, ,
	\end{equation}
where	
	\begin{equation}
	\mu (s) \equiv s \, \left[ \rho \, s \, \wt{J} (s) \right] ^{1/2} \, .
	\end{equation}
	Here $\wt{J} (s)$ is the creep compliance of a given viscoelastic model in the Laplace domain.
	 
From the boundary condition we can easily infer the formal solution of Eq.~(\ref{eq-motion-laplace}), \ie
	\begin{equation} \label{eq-sol-laplace}
	\wt{r} (s, x) = \wt{r} _{0} (s) \, \exp \left[ - x \, \mu (s) \right] \, .
	\end{equation} 
	
	For a singular model it is known (see \eg \cite{Mainardi_BOOK10}) that $\mu (s)$ will have an expansion in terms of decreasing powers of $s$, as $s$ goes to infinity. Thus, one could write this expansion as   
	\begin{equation} \label{eq-mu-as}
	\mu (s) \overset{s \to \infty}{\sim} \sum _{k=0} ^{\infty} b_{k} \, s^{1 - \beta _{k}} \, , \qquad 0 = \beta _{0} < \beta _{1} < \cdots
	\end{equation} 
	Now, it is important to define two quantities whose relevance will appear clear later in this discussion. Then, let us denote $\mu _{+} (s)$ the sum of the first $m+1$ ($m \in \N_{0}$) terms of Eq.~(\ref{eq-mu-as}) that depend on positive powers of $s$, \ie
	\begin{equation} \label{eq-mu-plus}
	\mu _{+} (s) \equiv \sum _{k=0} ^{m} b_{k} \, s^{1 - \beta _{k}} \, , \qquad \beta _{k} \leq 1 \, , \,\, k = 0, 1, \ldots, m \, .
	\end{equation}
 	Moreover, we denote with $\mu _{-} (s)$ the remainder of the series (\ref{eq-mu-as}). 
	Therefore, taking profit of these definitions, we can rewrite the solution in Eq.~(\ref{eq-sol-laplace}), for $s \to \infty$, as follows
	\begin{equation} \label{eq-sol-laplace-as}
	\wt{r} (s, x) 
	\overset{s \to \infty}{\sim} 
	\wt{r} _{0} (s) \, \exp \left[ - x \, \mu _{+} (s) \right] \, \wt{R} (s, x) \, , 
	\end{equation}
with
\begin{equation}
\wt{R} (s, x) \equiv \exp \left[ - x \, \mu _{-} (s) \right] \, .
\end{equation}
	
	Now, from the equation of motion in the Laplace domain (\ref{eq-motion-laplace}), it is clear that  the function $\wt{R} (s, x)$ is required to satisfy the following differential equation
	\begin{equation} \label{eq-L}
	\mathscr{O} \, \wt{R} (s, x) = \left\{ \frac{\partial^{2}}{\partial x^{2}} - 2 \, \mu _{+} (s) \, \frac{\partial}{\partial x} - [ \mu ^2 (s) - \mu _+ ^2 (s) ]  \right\} \, \wt{R} (s, x) = 0 \, , 
	\end{equation}
	together with the condition
	\begin{equation}
	\wt{R} (s, 0) = 1 \, .
	\end{equation}

If we now expand $\mu (s)$ and $\mu _{+} (s)$ in powers of $s$, as $s \to \infty$, in Eq.~(\ref{eq-L}) and then divide the whole equation for the term containing the highest power of $s$, we get a new differential operator $\mathbb{L}$. This is nothing but the rescaled version of $\mathscr{O}$ and it is featured by the following asymptotic expansion 
\begin{equation}
\mathbb{L} \overset{s \to \infty}{\sim} \sum _{i = 0} ^{\infty} \frac{1}{s ^{\nu _{i}}} \, \mathbb{L} _{i} \, , \qquad 0 = \nu _{0} < \nu _{1} < \cdots \, ,
\end{equation}
where
\begin{equation*}
	\begin{aligned}
	& \mathbb{L} _{0} = \frac{\partial}{\partial x} \, , \\
	& \mathbb{L} _{i} = p_{i} \, \frac{\partial ^{2}}{\partial x ^{2}} + q_{i} \, \frac{\partial}{\partial x} + r_{i} \, , \qquad i = 1, 2, 3 \ldots \, ,
	\end{aligned}
	\end{equation*}
and where the coefficients $p _{i}$, $q_{i}$ and $r_{i}$, together with the exponents $\nu _{i}$, are determined by the asymptotic expansion of $\mu (s)$ as $s \to \infty$.
	
	Now, one can easily observe that $\wt{R} (s, x)$ can be express as a (negative) power series
	\begin{equation}
	\wt{R} (s, x) \overset{s \to \infty}{\sim} \sum _{k=0} ^{\infty} v _{k} (x) \, s^{- \lambda _{k}} \, , \qquad 0 \leq \lambda _{0} < \lambda _{1} < \cdots \, ,
	\end{equation}
	for some functions $v_{k} (x)$. Moreover, we also have that the operator $\mathbb{L}$ matches the conditions of the Friedlander-Keller theorem (see \cite{Buchen-Mainardi 1975, IC-AG-FM-TWLV, FK}). Therefore, given this last remark, for $\mathbb{L}$ one can prove (see \cite{Buchen-Mainardi 1975, IC-AG-FM-TWLV}) that the functions $v_{k} (x)$ are actually given by
	\begin{equation} \label{eq-poly}
	v_k (x) = \sum _{\ell = 0} ^k A_{k, \ell} \, \frac{x^\ell}{\ell !} 
	\, ,
	\end{equation}
	where the coefficients $A_{k, \ell}$ are computed by means of the recursive system
	\begin{equation} \label{eq-system-A}
\begin{cases}
  A_{k, \ell} = \delta_{k \ell} \qquad \ell = 0 \, , \\
  A_{k, \ell} = - \ds \sum _{i} \left[ \sum _{J} \left( p_i A_{J, \, \ell+1} + q_i A_{J, \, \ell} + r_i A_{J, \, \ell-1} \right)\delta_{J, \, j(i,k)} \right] \qquad 1 \leq \ell \leq k \, , \\
  A_{k, \ell} = 0 \qquad \ell>k \, .
 \end{cases}
\end{equation}

Moreover, the function $j(i, k)$ is deduced by the condition $\nu _{i} + \lambda _{j} = \lambda _{k}$ and the coefficients $\lambda _{k}$ are such that
	\begin{equation}
	\lambda _{0} = 0 \, , \qquad \lambda _{k} = \sum _{i = 1} ^{\infty} m_{i} \, \nu _{i} \, ,
	\end{equation}
	with the $m_{i}$'s are positive integers. 
		
	As discussed in \cite{Buchen-Mainardi 1975} and extended and perfected in a quite general fashion in \cite{IC-AG-FM-TWLV}, if we now plug what we have found so far for $\wt{R} (s, x)$ into Eq.~(\ref{eq-sol-laplace-as}), we have that
	\begin{eqnarray}
	\wt{r} (s, x) 
	\overset{s \to \infty}{\sim} 
	\sum _{k=0} ^\infty v_k (x) \,
	\, \wt{r} _{0} (s) \, s^{- \lambda _k} \, 
	\exp \left[ - x \, \mu _{+} (s) \right] \, .
	\end{eqnarray}		  
	Then, defining
	\begin{equation} \label{eq-phi-laplace}
	\wt{\Phi} _k (s, x) \equiv \wt{r} _{0} (s) \, s^{- \lambda _k} \, 
	\exp \left[ - x \left( \mu _{+} (s) - \frac{s}{c} \right) \right] \, ,
	\end{equation}
we get\footnote{If $J_{0} = 0$ we have that $1/c = 0$.}
	\begin{equation} \label{eq-fin-laplace}
	\wt{r} (s, x) \overset{s \to \infty}{\sim}
	\exp \left[ - \frac{x \, s}{c} \right] \, 
	\sum _{k=0} ^\infty v_k (x) \, \wt{\Phi} _k (s, x) \, .
	\end{equation}
	For a mathematical discussion of the functions $\wt{\Phi} _k (s, x)$ we invite the interested reader to refer to \cite{Buchen-Mainardi 1975, Mainardi_BOOK10}. 
	
	Then, inverting this last result back to the time domain (see \cite{Buchen-Mainardi 1975, Mainardi_BOOK10}) we get the asymptotic expansion for $r(t,x)$ as $t \to (x/c)^+$, \ie
	\begin{equation} \label{eq-fin}
	r (t, x) \,\, \overset{t \to (x/c)^+}{\sim} \,\, 
	\sum _{k=0} ^\infty v_k (x) \, \Phi _k \left(t - \frac{x}{c}, x \right) \, ,
	\end{equation}
where the functions $v_k (x)$ are computed by means of Eq.~(\ref{eq-poly}) and (\ref{eq-system-A}) and the functions $\Phi _k (t, x)$ are the Laplace inverse of the functions $\wt{\Phi} _k (s, x)$ in Eq.~(\ref{eq-phi-laplace}).
	
	Clearly, this is just a summary of the technique. The interested reader can find a complete description of this algorithm in \cite{IC-AG-FM-TWLV}.
	
\section{Propagation of transient waves in a viscoelastic Bessel body} \label{Sec-3}
	In this section we discuss the problem of the propagation of transient pulses within a semi-infinite Bessel medium. Such a problem has a very well known formal solution in the Laplace domain (see \eg \cite{Mainardi_BOOK10}), however inverting back to the time domain is often very difficult even for simple viscoelastic models. 

	Here we present an asymptotic expansion of the solution of the propagation problem in the neighborhood of the pulse onset. To do this, we take profit of the Buchen-Mainardi algorithm, discussed and extended in Section \ref{Sec-2}.  For sake of simplicity, we consider two inputs: $r_0(t)=\theta (t)$ (or, in the Laplace domain, $\wt{r} _0(s)= 1/s$) and an inpulsive input $r_0(t)=\delta (t)$ (or, in the Laplace domain, $\wt{r} _0(s)= 1$), where $\theta$ and $\delta$ are respectively the Heaviside and Dirac generalized functions. This is done without loss of generality, indeed the general response to an arbitrary input $r_0 (t)$ can be obtained by convolution. 
	
	In order to further simplify our discussion, given that for a general Bessel body we have that $J_0 > 0$, let us also assume $J_0 = c = \rho = 1$. Then, as discussed in \cite{IC-AG-FM-2016, IC-AG-FM-Conf}, for a Bessel body we have that 
\begin{equation} \label{eq-bessel-creep}
s \, \widetilde{J} (s; \, \nu) = 1 + \wt{\Psi} (s; \nu) \overset{s\to \infty}{\sim} 1 + 2 \, (\nu + 1) \, s^{-1/2} \, .
\end{equation}
Hence,
\begin{equation} \label{eq-mu}
\begin{split}
\mu (s) &= s \, [s \, \widetilde{J} (s; \, \nu)]^{1/2} 
\sim s \, [1 + 2 \, ( \nu + 1) \, s^{-1/2}]^{1/2} \sim\\
&\sim s \, \left[ 1 + (\nu + 1) \, s^{-1/2} - \frac{(\nu + 1)^2}{2} \, s^{-1} + \sum _{n = 3} ^\infty  \binom {1/2} {n} \, 2^{n} \, (\nu + 1)^n \, s^{-n/2} \right] \sim\\
&\sim s + (\nu + 1) \, s^{1/2} - \frac{(\nu + 1)^2}{2} + 
\sum _{n = 3} ^\infty  \binom {1/2} {n} \, 2^{n} \, (\nu + 1)^n \, s^{1-n/2} \, ,
\end{split}
\end{equation}
as $s \to \infty$, where we have also taken profit of the Taylor expansion
$$ (1 + x)^\alpha \overset{x \ll 1}{=} \sum _{n = 0} ^\infty  \binom {\alpha} {n} \, x^n \, . $$	
	Following the procedure presented in Section \ref{Sec-2}, we set
\begin{eqnarray}
	\mu _+ (s) &\!\!=\!\!& s + (\nu + 1) \, s^{1/2} - \frac{(\nu + 1)^2}{2} \, , \label{eq-mu+} \\
	\mu _- (s) &\!\!=\!\!& \sum _{n = 3} ^\infty  \binom {1/2} {n} \, 2^{n} \, (\nu + 1)^n \, s^{1-n/2} \, . \label{eq-mu-} 
\end{eqnarray}	
Now, from Eq.~(\ref{eq-mu}) and Eq.~(\ref{eq-mu+}) we have
\begin{eqnarray}
\mu ^2 (s) &\!\!\overset{s\to \infty}{\sim}\!\!& s^2 \, [1 + 2 \, (\nu + 1) \, s^{-1/2}] 
\sim s^2 + 2 \, (\nu + 1) \, s^{3/2} \, , \\
\mu _+ ^2 (s) &\!\!=\!\!& \left( s + (\nu + 1) \, s^{1/2} - \frac{(\nu + 1)^2}{2} \right) ^2 =\\
&\!\!=\!\!& s^2 + 2 \, (\nu + 1) \, s^{3/2} - (\nu + 1)^3 \, s^{1/2} + \frac{(\nu + 1)^4}{4} \, , \notag
\end{eqnarray}
from which we get
\begin{equation}
\mu ^2 (s) - \mu _+ ^2 (s) \overset{s\to \infty}{\sim} (\nu + 1)^3 \, s^{1/2} - \frac{(\nu + 1)^4}{4} \, .
\end{equation}
Therefore, the differential operator $\mathbb{L}$, defined in Eq.~(\ref{eq-L}), for a Bessel medium reduces to
\begin{equation}
\begin{split}
\mathscr{O} \!\sim\! \Bigg\{ \frac{\partial^{2}}{\partial x^{2}} 
&- 2 \, \Bigg[ s + (\nu + 1) \, s^{1/2} - \frac{(\nu + 1)^2}{2} \Bigg] \, \frac{\partial}{\partial x}\\
&- \Bigg[ (\nu + 1)^3 \, s^{1/2} - \frac{(\nu + 1)^4}{4} \Bigg]\Bigg\} \, ,
\end{split}
\end{equation}
as $s \to \infty$.
If we rescale this operator by dividing the previous expression by the term containing the highest power of $s$, i.e. $-2 \, s$, we get
\begin{equation}
\begin{split}
\mathbb{L} \equiv \frac{\mathscr{O}}{-2 \, s} \!\sim\! \Bigg\{ - \frac{1}{2 \, s} \frac{\partial^{2}}{\partial x ^{2}} &+ 
\Bigg[ 1 + (\nu + 1) \, \frac{1}{s^{1/2}} - \frac{(\nu + 1)^2}{2} \frac{1}{s} \Bigg] \, \frac{\partial}{\partial x} \\
&+ \Bigg[ \frac{(\nu + 1)^3}{2} \frac{1}{s^{1/2}} - \frac{(\nu + 1)^4}{8} \frac{1}{s} \Bigg]\Bigg\} \, .
\end{split}
\end{equation}
Then, grouping the differential operators in the previous expression in terms of powers of $s$, we get
\begin{equation}
\begin{split}
\mathbb{L} \!\sim\! \Bigg\{ 
\frac{\partial}{\partial x} &+
\Bigg[ (\nu + 1) \, \frac{\partial}{\partial x} + \frac{( \nu + 1)^3}{2} \Bigg] \frac{1}{s^{1/2}} \\
&+
\Bigg[ - \frac{1}{2} \frac{\partial^{2}}{\partial x^{2}} - \frac{(\nu + 1)^2}{2} \frac{\partial}{\partial x} - \frac{(\nu + 1)^4}{8} \Bigg] \frac{1}{s} \, \Bigg\} \, .
\end{split}
\end{equation}
Therefore, the rescaled operator can be re-expressed as
\begin{equation}
\mathbb{L} \sim \mathbb{L} _{0} + \frac{1}{s^{1/2}} \, \mathbb{L} _{1} + \frac{1}{s} \, \mathbb{L} _{2} \, ,
\end{equation}
where,
\begin{eqnarray}
\mathbb{L} _{0} &=& \frac{\partial}{\partial x} \, , \\
\mathbb{L} _{1} &=& (\nu + 1) \frac{\partial}{\partial x} + \frac{(\nu + 1)^3}{2} \, , \\
\mathbb{L} _{2} &=& - \frac{1}{2} \frac{\partial ^{2}}{\partial x^{2}} - \frac{(\nu + 1)^2}{2} \frac{\partial}{\partial x} - \frac{(\nu + 1)^4}{8} \, .
\end{eqnarray}
Thus, comparing the rescaled operator with the first hypothesis of the theorem by Friedlander and Keller, \ie
$$ \mathbb{L} \overset{\varepsilon \to 0}{\sim} \sum _{i = 0} ^{\infty} \varepsilon ^{\nu _{i}} \, \mathbb{L} _{i} \, , \qquad 0 = \nu _{0} < \nu _{1} < \cdots \, , $$
we infer that for a Bessel solid $\varepsilon = 1/s$, $\nu _{0} = 0$, $\nu _{1} = 1/2$, $\nu _{2} = 1$ and $\mathbb{L} _i = 0$, $\forall i \geq 3$.

	Moreover, the coefficients $\lambda _{k}$ can be deduced by the system
\begin{equation} \label{eq-lambdak}
 \begin{aligned} 
& \lambda _0 =0 \,, \\
& \lambda _k= {\ds\sum_{i=1}^N m_i\, \nu _i}\,, \quad m_{i} \in \N \, , \,\, k= 1,2, \dots \, ,
\end{aligned}
\end{equation}
where $N$ is the number of terms in the asymptotic expansion of $\mathbb{L}$.

	Indeed, in our case we have that the only non-vanishing terms that appear in the expansion of the operator $\mathbb{L}$ feature the following values of $\nu _{i}$,
	$$ \nu_{0} = 0 \, , \quad \nu _{1} = \frac{1}{2} \, , \quad \nu _{2} = 1 \, , $$
	therefore, form Eq.~(\ref{eq-lambdak}) we have 
	\begin{equation}
	\lambda _{k} = \sum_{i=1}^\infty m_i\, \nu _i = \frac{m_{1}}{2} + m_{2} \in \N / 2
	\end{equation}
	from which we can infer that the coefficients $\lambda _{k}$ in the Buchen-Mainardi algorithm, for the scenario of our concern, are given by $\lambda _{k} = k/2$, with $k= 1,2, \dots$.

	Finally, form the condition $\nu _{i} + \lambda _{j} = \lambda _{k}$, we infer that $j (i, k) = k - 2 \, \nu _{i}$.

	Now, taking profit of the system in Eq.~(\ref{eq-system-A}), we can compute the values of the coefficients $A_{k, \ell}$ for a generic Bessel medium. In particular, given the previous discussion, we have that the system in Eq.~(\ref{eq-system-A}) can be rewritten as
\begin{equation}
 \begin{cases}
  A_{k, 0} = \delta_{k 0} \qquad \ell=0 \\
  A_{k, \ell} = - \ds \sum _{i=1,2} \sum _{J} \left( p_i A_{J,\ell+1} + q_i A_{J, \ell} + r_i A_{J, \ell-1} \right) \delta_{J, \, k - 2 \, \nu _{i}} \qquad 1 \leq \ell \leq k \\
  A_{k, \ell} = 0 \qquad \ell>k
 \end{cases}
\end{equation}
In particular, the second line can be rewritten, dropping the sum and the delta, as
\begin{equation}
A_{k, \ell} = - \ds \sum _{i=1} ^2 \Big( p_i A_{k - 2 \, \nu _{i}, \, \ell+1} + q_i A_{k - 2 \, \nu _{i}, \, \ell} + r_i A_{k - 2 \, \nu _{i}, \, \ell-1} \Big) \qquad 1 \leq \ell \leq k
\end{equation}
Now, accordingly with the discussion in Section \ref{Sec-2}, we can immediately read off the values of all $p_{i}$'s, $q_{i}$'s and $r_{i}$'s from the expressions for $\mathbb{L} _{1}$ and $\mathbb{L} _{2}$, indeed
\begin{equation}
\begin{split}
p_{1} &= 0 \, , \,\,\, q_1= \nu + 1 \, , \,\,\, r_1=\ds\frac{(\nu+1)^3}{2} \, ; \\
p_{2} &= -\ds\frac{1}{2} \, , \,\,\, q_2=-\ds\frac{(\nu+1)^2}{2} \, , \,\,\, r_2=-\ds\frac{(\nu+1)^4}{8} \, ; \\
\end{split}
\end{equation}
whereas $p_{i} = q_{i} = r_{i} = 0 \, , \,\, \forall i \geq 3$.

	Then, if we focus on the coefficients such that $1 \leq l \leq k$ we get
	\begin{equation}
	\begin{split}
	A_{k, \, \ell} &= - \ds \sum _{i=1} ^2 \Big( p_i A_{k - 2 \, \nu _{i}, \, \ell+1} + q_i A_{k - 2 \, \nu _{i}, \, \ell} + r_i A_{k - 2 \, \nu _{i}, \, \ell-1} \Big)=\\
	 &= - \Big[ \left( p_1 A_{k - 2 \, \nu _{1}, \,\ell+1} + q_1 A_{k - 2 \, \nu _{1}, \, \ell} +
	 r_1 A_{k - 2 \, \nu _{1}, \, \ell-1} \right) +\\
	 &+\left( p_2 A_{k - 2 \, \nu _{2}, \, \ell+1} + q_2 A_{k - 2 \, \nu _{2}, \, \ell} +
	 r_2 A_{k - 2 \, \nu _{2}, \, \ell-1} \right) \Big] =\\
	 &= - \Bigg[ \left(  (\nu + 1) \, A_{k - 1, \, \ell} +
	 \frac{(\nu+1)^3}{2} \, A_{k - 1, \, \ell-1} \right) +\\
	 &+\left( -\frac{1}{2} \, A_{k - 2, \, \ell+1} - \frac{(\nu+1)^2}{2} \, A_{k - 2, \, \ell} 
	 - \frac{(\nu+1)^4}{8} \, A_{k - 2, \, \ell-1} \right) \Bigg] \, .
	\end{split}
	\end{equation}
	Thus,
\begin{equation} \label{eq-Akl}
\begin{split}
A_{k, \,  \ell} &= -\frac{1}{2} \Bigg[ 2 \, (\nu +1) \, A_{k-1, \, \ell} + (\nu +1)^{3} \, A_{k-1, \, \ell-1} \\
& - A_{k-2, \, \ell+1} - (\nu + 1) ^{2} \, A_{k-2, \, \ell} - \frac{\left(\nu +1\right)^4}{4} \, A_{k-2, \, \ell-1} \Bigg] \,
\end{split}
\end{equation}

	Now, if we set $\nu = - 1/2$ these coefficients reduce to
\begin{equation}
 A_{k, \, \ell} = -\frac{1}{2} \Bigg[ A_{k-1, \, \ell} + \frac{1}{8} A_{k-1, \, \ell-1}- A_{k-2, \, \ell+1} - \frac{1}{4} A_{k-2, \, \ell} - \frac{1}{64} A_{k-2, \, \ell-1} \Bigg] \, ,
\end{equation}
that corresponds exactly to the one in \cite{Buchen-Mainardi 1975} for the \textit{Fractional Maxwell model of order 1/2}.

	In order to conclude this section, we still need to explicate all the parts of Eq.~(\ref{eq-fin}). First of all, let us rewrite Eq.~(\ref{eq-fin}) in terms of the polynomials $v_{k} (x)$, \ie
	\begin{equation}
	r (t, x) \,\, \overset{t \to (x/c)^+}{\sim} \,\, 
	\sum _{k=0} ^\infty \,  \sum _{\ell = 0} ^k A_{k, \ell} \, \frac{x^\ell}{\ell !} \, \Phi _k \left(t - x, x \right) \, .
	\end{equation}
	Then, to compute $\Phi _k \left(t - x, x \right)$ one just need to perform the inversion of its Laplace transform in Eq.~(\ref{eq-phi-laplace}). 
	
\subsection{Step Response} 
	If we consider a step input $\wt{r} _{0} (s) = 1/s$ we have that (see Appendix \ref{app-a})
	\begin{equation}
	\Phi _{k} (t - x, x) = \exp \left[ \frac{(\nu + 1)^2}{2} \, x \right] \, (t - x)^{k/2} \, F_{1/2} \left(z(\nu, t, x) , \, \frac{k}{2} \right) \, ,
	\end{equation}
	where
	$$ z(\nu, t, x) = \frac{(\nu +1) \, x}{(t - x) ^{1/2}} \, . $$
	Here we have also taken profit of the fact that for a Bessel body $\lambda _{k} = k/2$, as shown above, and that
	\begin{equation*}
	\mathcal{L} \left[ \theta (t - x) \, f( t - x) \, ; \, s \right] = e^{- x \, s} \, \wt{f} (s) \, ,
	\end{equation*}
	and
	\begin{equation*}
		F_{1/2} \left(z , \, \frac{k}{2} \right) = 2^{k} \mathscr{I}^{k} \texttt{erfc} \left( \frac{z}{2} \right) 
		\equiv \frac{2^{k+1}}{k! \, \sqrt{\pi}} \int _{z/2} ^{\infty} \left( u - \frac{z}{2} \right) ^{k} e^{-u^{2}} \, du \, .
	\end{equation*}

	Given the previous results, we are now capable to write down the explicit wave-front expansion, given a step input, for a generic viscoelastic Bessel medium, \ie
	\begin{equation} \label{eq-short-time}
	r (t, x) \sim 
	\exp \left[ \frac{(\nu + 1)^2}{2} \, x \right] \,
	\sum _{k=0} ^\infty \,  \sum _{\ell = 0} ^k A_{k, \ell} \, \frac{x^\ell}{\ell !}  \, 
	(t - x)^{k/2} \, F_{1/2} \left(z(\nu, t, x) , \, \frac{k}{2} \right) 	\, .
	\end{equation}
	where $z(\nu, t, x)$ is defined as above and where the coefficients $A_{k, \ell}$ are determined as in Eq.~(\ref{eq-Akl}).

\begin{figure}[h!] 
\centering
\includegraphics[width=10cm]{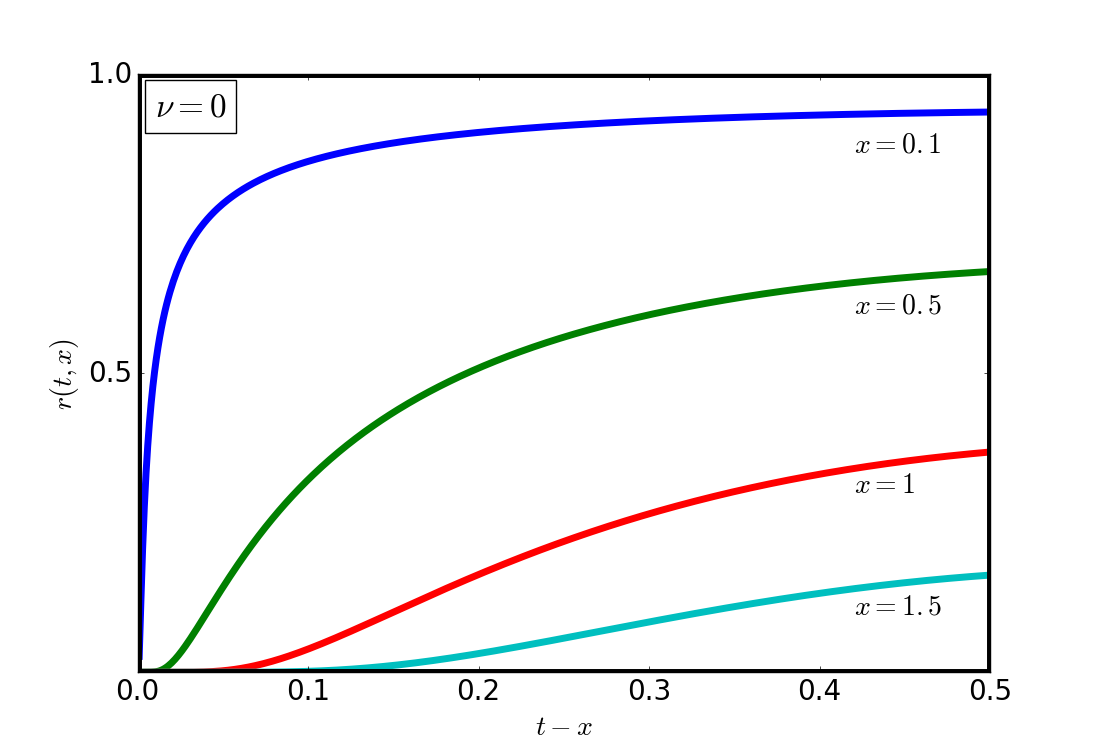}
\caption{The step response for the Bessel model of order $\nu=0.0$ depicted versus $t - x$ for some fixed values of $x$.\label{Fig-1}}
\end{figure}

	In Fig.~\ref{Fig-1} we show some plots of the wave-front expansion provided in Eq.~(\ref{eq-short-time}). We consider the specific case of $\nu = 0$ because it will tourn out to be relevant in biophysics.

\subsection{Impulse Response}
	Analogously, if we consider impulsive input $r_0 (t) = \delta (t)$, \ie $\wt{r} _0 (s) = 1$, one can easily show that
\begin{equation}
	\Phi _{k} (t - x, x) = \exp \left[ \frac{(\nu + 1)^2}{2} \, x \right] \, (t - x)^{(k-2)/2} \, F_{1/2} \left(z(\nu, t, x) , \, \frac{k - 2}{2} \right) \, ,
	\end{equation}
	where
	$$ z(\nu, t, x) = \frac{(\nu +1) \, x}{(t - x) ^{1/2}} \, . $$
Therefore, the wave-front expansion for a Bessel model of order $\nu$, given an impulsive input, is provided by
\begin{equation} \label{eq-short-time-2}
	r (t, x) \sim 
	\exp \left[ \frac{(\nu + 1)^2}{2} \, x \right] \,
	\sum _{k=0} ^\infty \,  \sum _{\ell = 0} ^k A_{k, \ell} \, \frac{x^\ell}{\ell !}  \, 
	(t - x)^{(k-2)/2} \, F_{1/2} \left(z(\nu, t, x) , \, \frac{k - 2}{2} \right) 	\, .
	\end{equation}
	where $z(\nu, t, x)$ is defined as above and where the coefficients $A_{k, \ell}$ are determined as in Eq.~(\ref{eq-Akl}). 

\begin{figure}[h!]
\centering
\includegraphics[width=10cm]{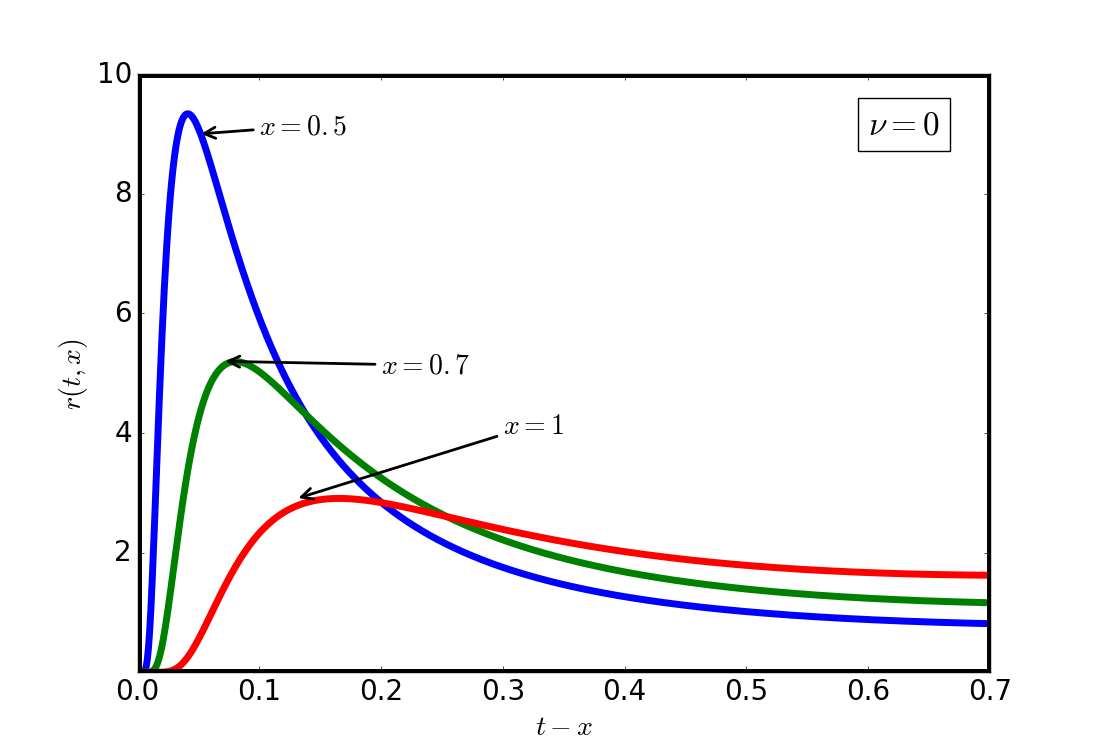}
\caption{The impulse response for the Bessel model of order $\nu=0.0$ depicted versus $t - x$ for some fixed values of $x$.} \label{Fig-2}
\end{figure}

In Fig.~\ref{Fig-2} we show some plots of the wave-front expansion provided in Eq.~(\ref{eq-short-time-2}).

	The case $\nu = 0$ corresponds to a peculiar model for large arteries \cite{AG-FM_MECC16, AG-FCAA-2017}, in which they can be thought of as fluid-filled elastic tubes. Therefore, the discussion concerning the impulse response appears to be particularly relevant. Indeed, an impulsive boundary condition represents a first rough approximation for the muscular pulse sent by the heart through arteries. Clearly, this is not an exact model for such kind of impulses given that the aortic valve, for example, would not resist an ``infinite stress'' without breaking. Nonetheless, the time evolution of the system after the pulse is basically the same. 
	
	Now, in Fig.~\ref{Fig-2} we show the impulsive response function as seen by a static ``observer'' placed in a given position $x$ of an artery. This plot clearly tells us that, at $x$, the signal will reach its (finite!) peak at 
$$t = \frac{x}{c} + \Delta $$
where $\Delta > 0$ is a time delay due to the viscosity of the fluid. So, we have that the blood viscosity slows down the motion of the pulse and dissipates part of its energy, leading to a damping of its peak. Moreover, the further away we put our observer from the edge of the artery, the stronger is the dumping of the peak of the pulse.

\section{On the long time behavior of transient waves in a Bessel solid} \label{Sec-4} 
	It is quite known in the theory of Laplace transform that, under fairly general conditions, the asymptotic behavior in the time domain is strictly related to the asymptotic behavior in the Laplace domain. In particular, let $f(t)$ be a suitable function in the time domain. Then, denoting with $\wt{f} (s)$ the Laplace transform of $f(t)$, the asymptotic behavior of $f(t)$ as $t \to \infty$ is uniquely determined by the asymptotic expansion of $\wt{f} (s)$ as $s \to 0$. This is basically the content of the first Tauberian theorem\footnote{Notice that in Section \ref{Sec-2}, in order to perform the wave-front expansion, we took profit the second Tauberian theorem, \ie the one that relates the asymptotic behavior of $f(t)$ as $t \to \infty$ to the asymptotic expansion of $\wt{f} (s)$ as $s \to 0$.}. For further informations about physical applications of this technique, see \eg \cite{IC-AG-FM-2016, AG-FM_MECC16, AG-FM-EPJP, AG-FCAA-2017, Mainardi_BOOK10}.

	Therefore, the long time asymptotic behavior of our response function $r(t, x)$ is formally equivalent to the asymptotic behavior of $\wt{r} (s, x)$ as $s \to 0$. Hence, recalling the formal solution of the equation of motion (\ref{eq-motion-laplace}), in the Laplace domain, \ie
	$$ \wt{r} (s, x) = \wt{r} _{0} (s) \, \exp \left[ - x \, \mu (s) \right] \, , $$
in order to infer the long time behavior of $r(t, x)$ we just need to compute the expansion for $\mu (s)$, as $s \to 0$, and then invert back to the time domain.

\subsection{Step Response}	
	First, let us consider the case of a step input $\wt{r} _{0} (s) = 1/s$. 
	
	For a body whose viscoelastic description is modeled in terms of Bessel models \cite{IC-AG-FM-2016} we have
	\begin{equation}
	s \, \widetilde{J} (s, \nu) = 1 + \frac{2(\nu + 1)}{\sqrt{s}} \frac{I_{\nu + 1}(\sqrt{s})}{I_{\nu}(\sqrt{s})} \overset{s\to0}{=} 1 + \frac{4 (\nu +1) (\nu +2)}{s} + \frac{\nu + 1}{\nu + 3} + o(s) \, .
	\end{equation}
Thus,
\begin{equation}
\mu (s) = s \, [s \, \widetilde{J} (s, \nu)]^{1/2} \overset{s\to0}{=} \sqrt{4 (\nu +1)(\nu+2)} \, s^{1/2} \left[ 1 + \frac{1}{4 (\nu +1)(\nu+3)} \, s \right] + o (s^{5/2}) \, .
\end{equation}
Then, we can compute the behavior of the solution of the wave equation for our viscoelastic model as $s \to 0$, i.e. $t \to +\infty$, indeed
\begin{equation} \label{eq-r}
\begin{split}
\widetilde{r} (s, x) &= \widetilde{r} _{0} (s) \, \exp \left[ - x \, \mu (s) \right] = \frac{1}{s} \, \exp \left[ - x \, \mu (s) \right] \overset{s\to0}{\sim} \\
&\sim \frac{1}{s} \, \exp \left[ - \sqrt{4 (\nu +1)(\nu+2) \, x^{2}} \, s^{1/2} \right] \, .
\end{split}
\end{equation} 
Now, taking profit of the fact that (see \cite{AS 1965})
\begin{equation}
\mathcal{L} \left[ \texttt{erfc} \left( \frac{1}{2} \sqrt{\frac{a}{t}} \right)\, ; \,\, s \right] = 
\frac{1}{s} \, \exp \left( - \sqrt{a s} \right) \, ,
\end{equation} 
where $\mathcal{L} [f(t) \, ; \, s] \equiv \wt{f} (s)$, then from Eq.~(\ref{eq-r}) we immediately get
\begin{equation}
r (t, x) \sim \texttt{erfc} \left[ \sqrt{(\nu +1)(\nu+2)} \, \frac{x}{\sqrt{t}} \right] \quad \mbox{as} \,\,\, t \to + \infty \, .
\end{equation} 

\begin{figure}[h!]
\centering
\includegraphics[width=9cm]{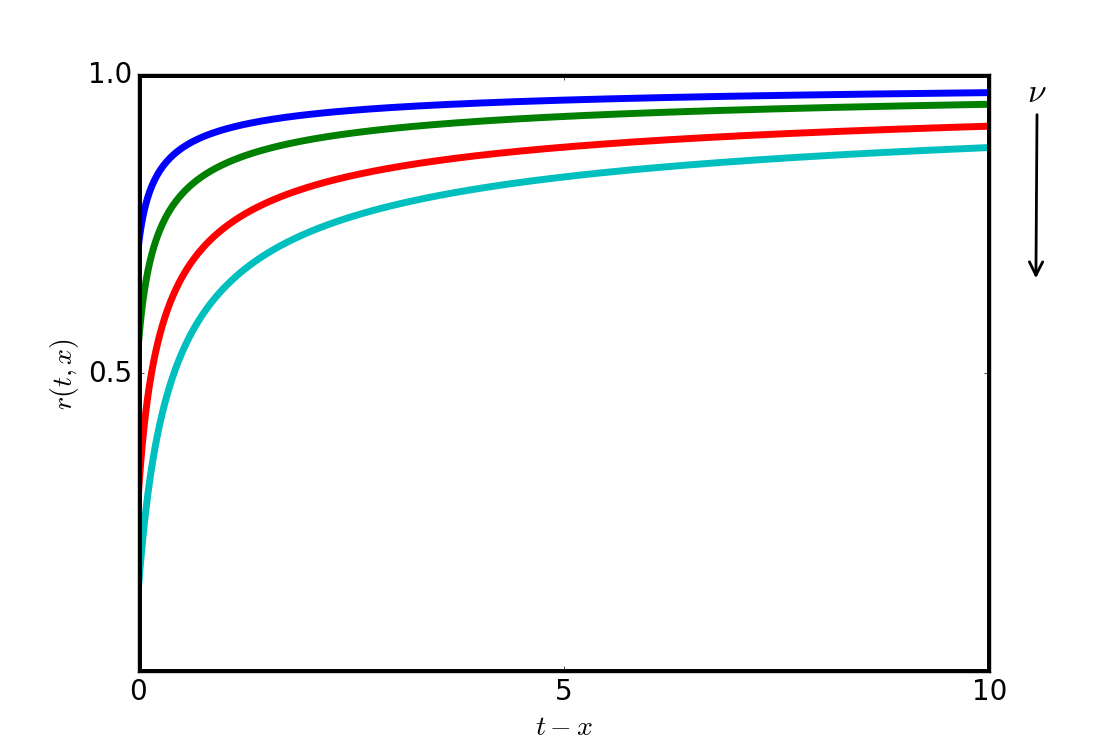}
\caption{The long time asymptotic expansion of the step response for $\nu = -0.5, 0.0, 1.0, 2$ depicted versus $t - x$, for $x = 0.1$.}
\end{figure}

\subsection{Impulse Response} On the other hand, if we consider an impulsive input $r_0 (t) = \delta (t)$, \ie $\wt{r} _0 (s) = 1$, then Eq.~(\ref{eq-r}) reads
\begin{equation} \label{eq-r-2}
\begin{split}
\widetilde{r} (s, x) &= \widetilde{r} _{0} (s) \, \exp \left[ - x \, \mu (s) \right] = \exp \left[ - x \, \mu (s) \right] \overset{s\to0}{\sim} \\
&\sim \exp \left[ - \sqrt{4 (\nu +1)(\nu+2) \, x^{2}} \, s^{1/2} \right] \, .
\end{split}
\end{equation} 

	Then, recalling that (see \cite{AS 1965})
\begin{equation}
\mathcal{L} \left[ \frac{k}{2 \sqrt{\pi \, t^3}} \, \exp \left( - \frac{k^2}{4 \sqrt{t}} \right) \, ; \,\, s \right] = 
\exp \left( - k \, \sqrt{s} \right) \, ,
\end{equation}
we can invert back to the time domain the expression in Eq.~(\ref{eq-r-2}), getting
\begin{equation}
r (t, x) \sim 
\sqrt{\frac{(\nu +1)(\nu+2)}{\pi}} \frac{x}{t^{3/2}} \, 
\exp \left[ - \frac{(\nu +1)(\nu+2) \, x^{2}}{\sqrt{t}} \right]
\quad \mbox{as} \,\,\, t \to + \infty \, .
\end{equation}

\begin{figure}[h!]
\centering
\includegraphics[width=9cm]{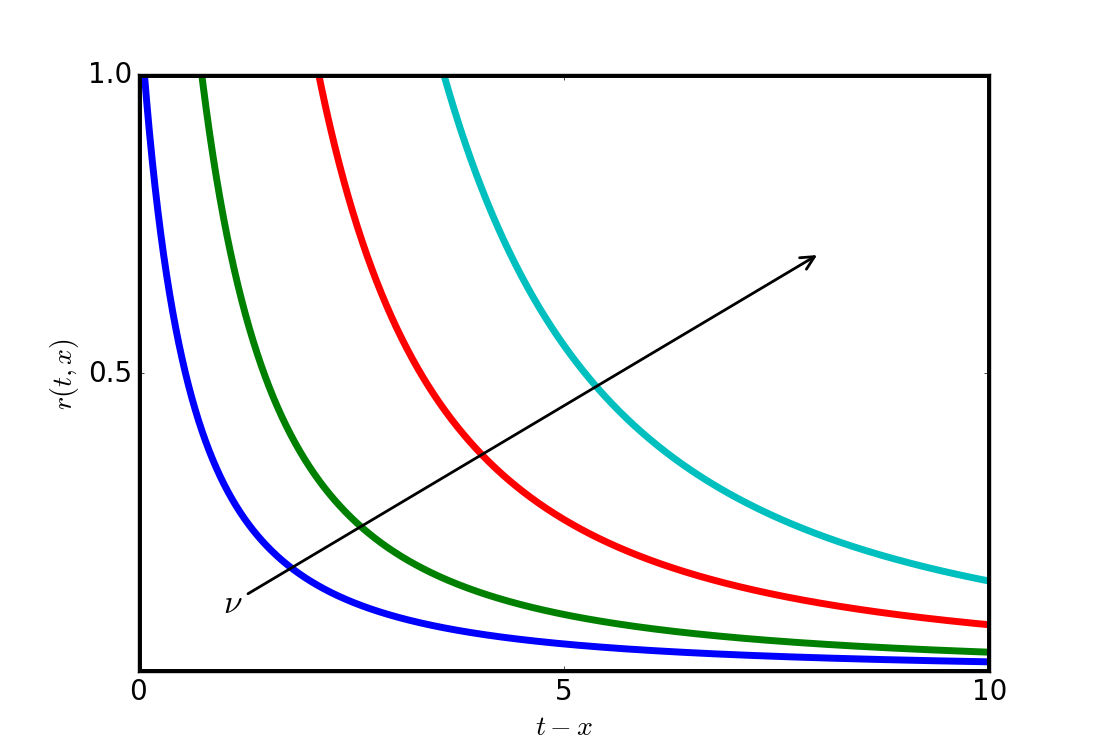}
\caption{The long time asymptotic expansion of the impulse response for $\nu = -0.5, 0.0, 1.0, 2$ depicted versus $t - x$, for $x = 1$.}
\end{figure}

\section{Conclusions}
	In this paper we discussed the propagation of transient waves in a semi-infinite linear viscoelastic medium described by a Bessel model of order $\nu$.
	
	In particular, in Section~\ref{Sec-2} we revisited the Buchen-Mainardi algorithm for the wave-front expansion for singular viscoelastic models.
	
	Then, in Section~\ref{Sec-3} we applied the previous method in order to compute the wave-front expansion for a Bessel body of order $\nu$, given two typical input conditions. First, we considered the ``canonical'' step input, in analogy with the seminal paper \cite{Buchen-Mainardi 1975} by Buchen and Mainardi. This input is particularly relevant because the singular behavior of the memory function at $t=0^+$ induces a wave-front smoothing of any initial discontinuity. For further details on this smoothing effect found in transient waves propagating in media with singular memory, we refer to Hanyga \cite{Hanyga}.  
	
	Then, we analyzed an impulsive boundary condition in order to provide some general remarks concerning a peculiar hemodynamical model, first proposed by Giusti and Mainardi in \cite{AG-FM_MECC16}. The result is that this model appears to be consistent with the expected behavior of a pressure wave as it propagates within a large artery.    

	Finally, in Section~\ref{Sec-4} we provided an analytic asymptotic description of the long time behavior of the response function for both the analyzed scenarios.

\section*{Acknowledgments}
	The work of the authors has been carried out in the framework of the activities of the National Group of Mathematical Physics (GNFM, INdAM). The authors would also like to thank V. A. Diaz and C. De Tommasi for valuable comments and discussions.


\appendix

\section{Useful Calculations} \label{app-a}
	As we discussed in Section \ref{Sec-2}, in general we have that (setting $c=1$)
	\begin{equation}
	\wt{\Phi} _k (s, x) \equiv \wt{r} _{0} (s) \, s^{- \lambda _k} \, 
	\exp \left[ - x \left( \mu _{+} (s) - s \right) \right] \, .
	\end{equation}
	If we consider, in particular, the case in which $\wt{r} _{0} (s) = 1/s$, the previous expression can be rewritten as
	\begin{equation}
	\wt{\Phi} _k (s, x) \equiv s^{- (\lambda _k + 1)} \, 
	\exp \left[ - x \left( \mu _{+} (s) - s \right) \right] \, .
	\end{equation}
	Now, for a Bessel solid we have
	\begin{equation}
	\mu _+ (s) = s + (\nu + 1) \, s^{1/2} - \frac{(\nu + 1)^2}{2} \, ,
	\end{equation}
	thus,
	\begin{equation} \label{eq-problem-A}
	\wt{\Phi} _k (s, x) \equiv \exp \left[ \frac{(\nu + 1)^2}{2} \, x \right] \, s^{- (\lambda _k + 1)} \, 
	\exp \left[ - (\nu + 1) \, x \, s^{1/2} \right] \, .
	\end{equation}
	If we then consider a function $\wt{f} (s)$, in the Laplace domain, defined as
	\begin{equation} \label{eq-1A}
	\wt{f} (s) = s^{-(\alpha + 1)} \, e^{- \beta \, s ^{\gamma}} \, ,
	\end{equation}
	and we denote with
	$$ \mathcal{E} _{\gamma} (\beta , t) \equiv
	 \mathcal{L} ^{-1} \left[ e^{- \beta \, s ^{\gamma}} \, ; \, t \right] \, , $$
	one can immediately infer that
	\begin{equation} \label{eq-2A}
	f(t) = \frac{t^{\alpha}}{\Gamma (\alpha + 1)} \, \ast \, \mathcal{E} _{\gamma} (\beta , t) \, .
	\end{equation}
	Moreover, expanding the exponential in Eq.~(\ref{eq-1A}) and then inverting term by term, one can easily prove that
	\begin{equation}
	\mathcal{E} _{\gamma} (\beta , t) = \sum _{k = 0} ^{\infty} \frac{(-\beta) ^{k} \, t^{- (\gamma \, k + 1)}}{k! \,\, \Gamma (- \gamma \, k)} \, , \qquad t > 0 \, .
	\end{equation}
 	
	Now, as suggested in \cite{Buchen-Mainardi 1975}, the easiest manner to compute the inverse of the wave-front expansion is thought recurrence relations. Therefore, for this purpose we can introduce the following function:
	\begin{equation}
	F_{\gamma} (z, \alpha) = 
	t^{- \alpha} \left[ \frac{t^{\alpha}}{\Gamma (\alpha + 1)} 
	\, \ast \, 
	\mathcal{E} _{\gamma} (\beta , t)\right] \, , \quad z (\beta, \gamma, t) 
	= \frac{\beta}{t^{\gamma}} \, ,
	\end{equation}   
	that satisfies the following recurrence relation
	\begin{equation}
	\alpha \, F_{\gamma} (z, \alpha) 
	= 
	- \gamma \, z \, F_{\gamma} (z, \alpha - \gamma) + F_{\gamma} (z, \alpha - 1) \, .
	\end{equation}
	
	Therefore, the inversion of $\wt{\Phi} _k (s, x)$ in Eq.~(\ref{eq-problem-A}) is simply given by,
	\begin{equation}
	\Phi _{k} (t, x) = \exp \left[ \frac{(\nu + 1)^2}{2} \, x \right] \, t^{\lambda _{k}} \, F_{1/2} \left(z(\nu, t, x) , \, \lambda _{k} \right) \, ,
	\end{equation}
	where
	$$ z(\nu, t, x) = \frac{(\nu +1) \, x}{t^{1/2}} \, . $$


\end{document}